# Integrated RF MEMS/CMOS Devices

R. R. Mansour, S. Fouladi and M. Bakeri-Kassem
University of Waterloo
Waterloo, Ontario, Canada

*Abstract* - A maskless post-processing technique for CMOS chips is developed that enables the fabrication of RF MEMS parallel-plate capacitors with a high quality factor and a very compact size. Simulations and measured results are presented for several MEMS/CMOS capacitors. A 2-pole coupled line tunable bandpass filter with a center frequency of 9.5 GHz is designed, fabricated and tested. A tuning range of 17% is achieved using integrated variable MEMS/CMOS capacitors with a quality factor exceeding 20. The tunable filter occupies a chip area of 1.2 × 2.1 mm$^2$.

## I. INTRODUCTION

There is a need for intelligent reconfigurable radio Frequency (RF) front-ends that can achieve maximum hardware sharing for various standards. Such an RF front-end requires reconfigurable blocks rather than switched architectures. These reconfigurable blocks need to meet several RF specifications such as smaller size, lower loss, higher tuning range, linearity, and low power consumption. The continuous shrinkage of minimum feature size in current silicon-based technologies such as CMOS provides an opportunity to meet these demands. Moreover, using these technologies, it is possible to integrate the reconfigurable RF blocks with the core signal processing circuitry leading to the implementation of compact intelligent reconfigurable RF front-ends.

Tunable filters and amplifier tunable matching networks are among the reconfigurable blocks that if integrated with RF CMOS electronics can reduce the overall system size, weight and cost. A number of integrated tunable filters have been demonstrated using solid-state tuning elements. Despite the small area and high quality factor of these active tuning elements, the major disadvantage is their susceptibility to noise and lower power handling due to nonlinearities in the active semiconductor devices.

This paper demonstrates the feasibility of integrating MEMS with CMOS to realize reconfigurable front end with a superior RF performance in terms of linearity and insertion loss. It also presents a novel post-processing technique to create MEMS varactors in CMOS chips. The paper will also address other applications that can benefit from the successful integration of MEMS with CMOS.

## II. FABRICATION PROCESS

A CMOS micromachining process was reported in [1] to fabricate CMOS-MEMS tunable capacitors. The capacitors were based on interdigitated beam structure, and electro-thermal actuators were used for tuning. We propose a post-processing technique that enables the fabrication of parallel-plate tunable capacitors with a higher capacitance value, smaller size and zero dc power consumption using electrostatic actuation mechanism. The CMOS-MEMS post-processing of tunable capacitors starts with a chip fabricated using the TSMC 0.35μm CMOS technology. There are two polysilicon and four metal layers available through this process. The process offers the possibility for the fabrication of MEMS structures using metal interconnect layers. The layout of the tunable parallel-plate capacitor and its cross-sectional view are presented in Fig. 1. The top and bottom plates are made of M3 and M1 layers, respectively and the second metal layer (M2) is used as a sacrificial layer which will be removed after post processing in order to create an air gap between the plates. By using this approach after release, the total gap between the plates consists of two dielectric layers each having a thickness of 1μm and a 0.6μm air gap. The dielectric layer prevents the occurrence of a short circuit when the dc tuning voltage exceeds the pull-in voltage, and the top plate snaps down on the bottom plate.

Three maskless post-processing steps are required to construct the tunable capacitors. They also include improving RF performance by etching away the silicon under the device and eliminating substrate loss. A schematic view of the post-processing steps is presented in Fig. 2. The first processing stage involves the removal of silicon dioxide around the MEMS structure as presented in Fig. 2(b). This can be done using the last metal layer (M4) as a mask and reactive ion etching (RIE) of the oxide layer [3]. The main purpose of this step is twofold. First, to expose the sacrificial M2 layer that will be etched during the next post-processing step. Secondly, to create windows through the oxide down to the silicon substrate, which are used to form a trench under the capacitors improving their quality factor and self-resonance frequency. It is important to keep an oxide layer around the structural metal layers (M1 and M3) to protect them during the removal of the M2 sacrificial layer. This is accomplished by extending the masking metal layer (M4) on top of the structural layers. However, the second metal layer which is required to be exposed after the RIE step, should be extended further than M4.

As shown in the cross-sectional view of the proposed capacitor in Fig. 1, the extension of M4 over M1 and M3 is 2μm. assuming that the RIE etch recipe is anisotropic, this leaves a 2µm of side-wall oxide to protect the structural metal layers. The M4 layer also covers the CMOS electronics circuitry. Fig. 3(a) presents an SEM image of the capacitor after the first post-processing step. As shown in this figure, M2 and the silicon substrate are exposed and ready to be etched during the next step. The second post-processing step as shown in Fig. 2(b) involves the etching of the sacrificial layer and substrate using wet etching techniques. The sacrificial layer is removed using PAN etch followed by TiW etchant. In addition to the sacrificial layer removal, the exposed M4 layer used as a mask during the RIE step will be removed. For wet etching of silicon, diluted electronic grade TMAH is used. For an acceptable RF performance, a minimum trench depth of 75µm is required. An SEM image of the CMOS-MEMS capacitor after the wet etching step is shown in Fig. 3(b). As shown in this figure, in order to facilitate the etching of the sacrificial metal layer sufficient number of release holes are included on the top plate of the capacitor. During the last post-processing step, devices are rinsed in IPA and released in a critical point dryer (CPD) system to avoid any stiction problems. This is followed by a final RIE step that is required to remove the protecting oxide layer on top of M3 and RF pads as shown in Fig. 2(c). We have been also successful in implementing the process in 0.18 µm CMOS.

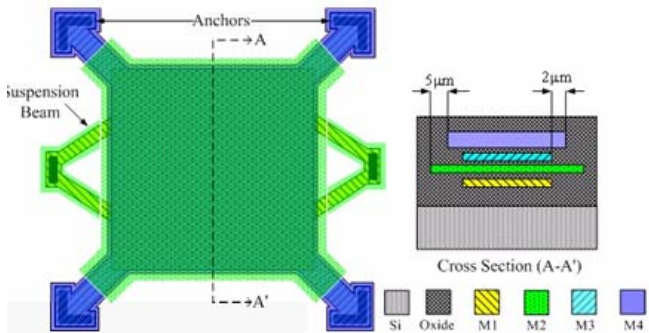

Figure 1 Layout and cross-sectional view of the parallel-plate MEMS tuning elements fabricated using the 0.35µm CMOS process.

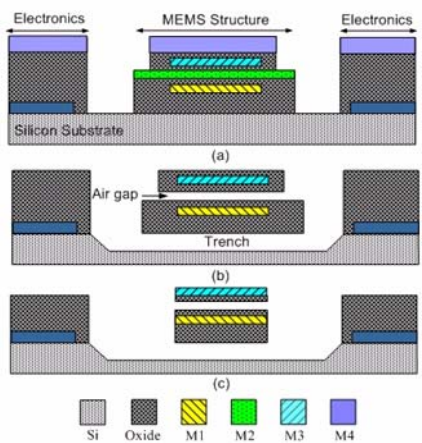

Fig. 2. CMOS-MEMS post-processing steps, (a) RIE oxide removal, (b)wet etching, CPD and (c).2nd oxide etching

Figure 4 shows an integrated 2-pole tunable filter implemented using four MEMS/CMOS varactors. The measured performance of the tunable filter is shown in the same figure [2]. Figure 5 illustrates SEM pictures of MEMS varactors implemented in CMOS chips using the above proposed post–processing technique. The varactor exhibits a Q value of better than 300 at 1.5 GHz [3].

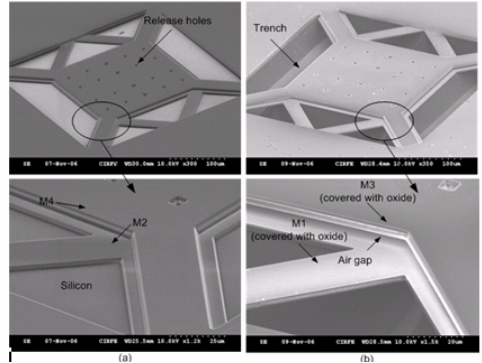

Fig. 3. SEM image of the capacitor after first (a) and second (b) post-processing steps.

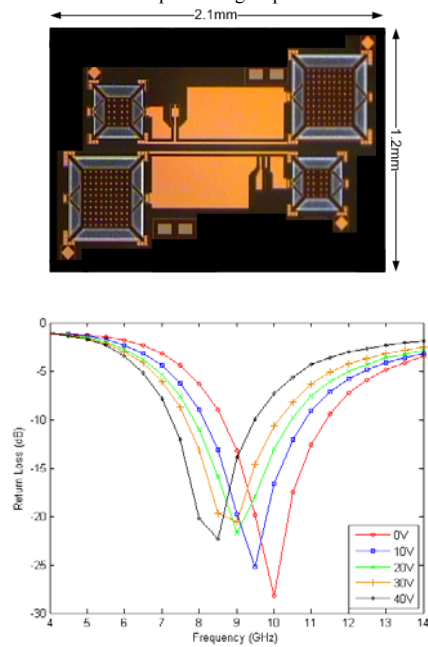

Figure 4. a) Optical photograph of the MEMC/CMOS tunable filter, b) the measured return loss at different bias voltages.

REFERENCES

[1] G. K. Fedder, and T. Mukherjee, "Tunable RF and Analog Circuits Using On-Chip MEMS Passive Components," *2005 IEEE Solid-State Circuits Conf. Dig.*, vol. 1, pp. 390-391, February 2005.

[2] S. Fouladi, M. Baker-Kassem, and R. R. Mansour," An Integrated Tunable Band-Pass Filter Using MEMS Parallel-Plate Variable Capacitors Implemented with 0.35 µm CMOS Technology, IEEE–International Microwave Symposium, IMS, pp.505-508, June 2007.

[3] M. Baker-Kassem, S. Fouladi, and R. R. Mansour," Novel High-Q MEMS Curled-plate Variable Capacitor Fabricated in 0.35 µm CMOS Technology," IEEE Transactions on Microwave Theory and Techniques, Vol 56, pp. 530-541, February 2008.